# Cyber Crossroads
## A Global Research Collaborative on Cyber Risk Governance

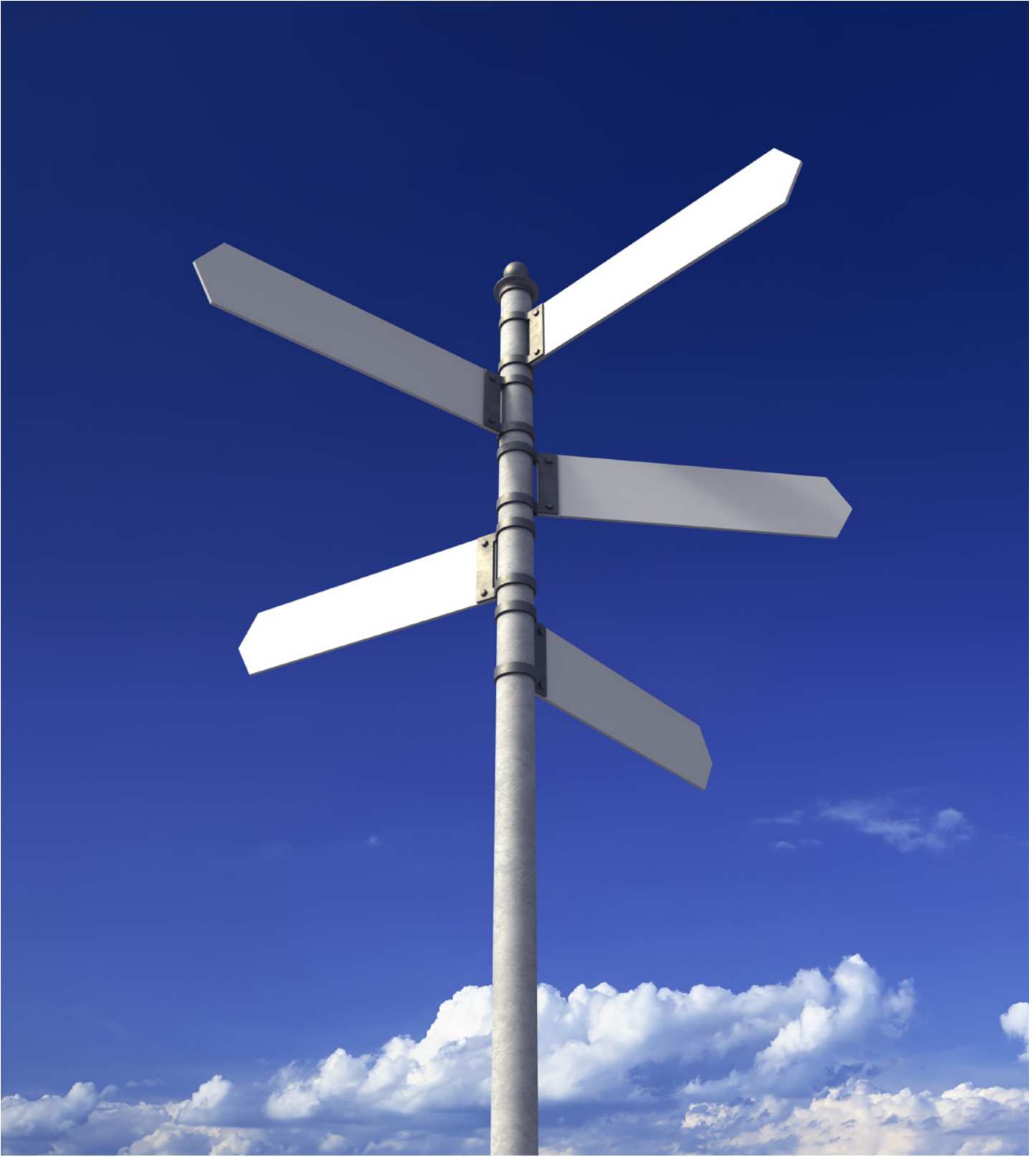

# Cybersecurity is at a Crossroads

Spending on cybersecurity products and services is expected to top $123 billion for 2020, more than double the $55 billion spent in 2011.[1] In that same period, cyber breaches quadrupled.[2] Organizations globally face increasing liabilities, while boards of directors grapple with a seemingly Sisyphean challenge. Cyber Crossroads was born out of these alarming trends and a realization that the world cannot go on funneling finite resources into an indefinite, intractable problem.

Cyber Crossroads brings together expertise from across the world, spanning aspects of the cyber problem (including technology, legal, risk, and economic) with the goal of creating a Cyber Standard of Care built through a global, not-for-profit research collaborative with no commercial interests. A Cyber Standard of Care should be applicable across industries and regardless of the organization size. It should be practical and implementable, with no requirement to purchase any product/service. Cyber Standard of Care should be woven into the existing governance fabric of the organization and it should not be yet another technical checklist, but a process/governance framework that can stand over time.

To achieve this, we engaged with cyber risk experts and practitioners with a variety of relevant expertise, secured the advice/guidance of regulators and legal experts across jurisdictions, and interviewed leaders from 56 organizations globally to understand their challenges and identify best practices.

Given that the risk and insurance industry is a thought leader in cyber risk, we secured sponsorship from leading companies in the sector, which also provided strategic guidance. In order to maintain research integrity, we did not incorporate sponsor's perspectives into the research; however, the research would not have been successful without their guidance and support.

We want to thank all members of the research team and all of the interlocutors who gave their valuable time and guidance. We hope this Cyber Crossroads report initiates the next phase of the world's cyber risk journey and provides a foundation for future research on managing cyber risk.

[1] Gartner | [2] Verizon Data Breach Report

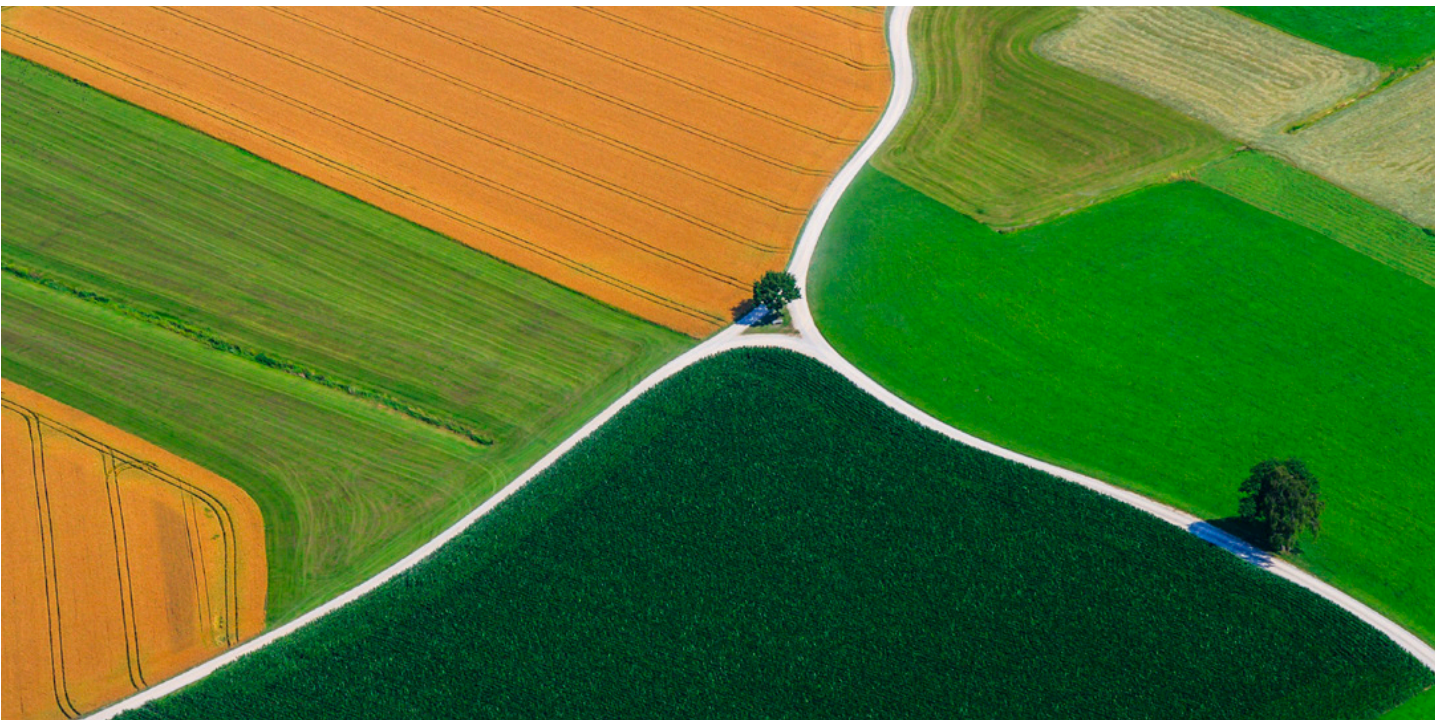

# Contents

# Imagine a patient presenting to a doctor with an unusual set of symptoms.

The doctor recalls a similar case from many years ago, but knows that treatment protocols have evolved. They consult the literature, speak with an expert, and begin a course of treatment reflecting the changes. Sadly, the patient fails to respond, and dies.

A review board finds the doctor met the standard of care, and takes no disciplinary action. In fact, the patient's family briefly discusses a lawsuit, but their attorney sees that the doctor met the medical standard of care, and does not sue.

Now imagine an organization under cyber-attack from a threat actor wielding ransomware. A single laptop is infected, and the employee reports to their manager, who suggests rebooting. Soon a PC in the office displays the same ransom note, and then another before someone contacts local IT support. A similar scene plays out in one of the company's offices in another state, and then another. It takes a few days for the organization to connect all the dots. By then, the damage is severe, and the company's share price takes a hit.

The lawsuits that follow zero in on the lack of coordinated planning, the slow response time, and the board of directors' inadequate oversight. The reputational damage is a boon to competitors.

What if there had been a Cyber Standard of Care for the company to follow, in much the same way a doctor abides by a medical standard of care? Just like with the doctor, there is no guarantee the company would have thwarted the attack, but there is a chance it could have minimized damage, established a clear rationale — and possibly a legal defense — for its response, and helped limit damage to its reputation.

Put simply, a Cyber Standard of Care is a governance framework for organizations to help them manage their cyber risk. At the moment, such a standard of care simply does not exist.

In this paper, Cyber Crossroads proposes a foundational Cyber Standard of Care inspired by the development of such standards in other disciplines, particularly medicine. We offer a systematic approach to ensure that any organization — across industries and size classes — can meet reasonable cybersecurity standards. It is important to be clear what we mean by cyber risk and cybersecurity. Our focus is the vulnerability of organizations to malicious attacks and employee errors that in one way or another threaten to damage their business. We are not concerned with day-to-day operational issues such as the stability, administration, and maintenance of systems.

Cyber Crossroads and the Cyber Standard of Care are the product of extensive collaboration involving world-renowned cybersecurity experts, corporate leaders, legal advisors, national regulators, and global insurers. Reflected throughout this report are the insights from interviews we conducted with senior executives at 56 organizations globally.

We hope this guide can help generate discussion and accelerate the cyber risk governance journey for all organizations.

# The Need for the Cyber Standard of Care

Current efforts to manage cyber risk are not keeping pace with the growing cyber threat. Despite organizations spending increasing amounts on cybersecurity products and services, cyber-attacks and resulting losses are becoming more frequent and severe.

Managers and organizational leaders responsible for cybersecurity lack an objective standard of care that offers guidance on how they should discharge their obligations. They would benefit from a clear standard for responsible cybersecurity management to which shareholders, customers, employees, regulators, courts, and even the court of public opinion could refer to evaluate their performance.

Directors and officers who are not specialists in cyber risk need clearer guidelines so they can marshal appropriate personnel, identify critical cyber risk factors, and allocate institutional resources effectively. Likewise, regulators and investors need a standard of care to use as they evaluate the performance of directors and officers.

In many commercial practice areas, when risks cannot be fully eliminated it is standard procedure to establish foundational principles and processes to guide risk management and mitigation. However, this has yet to be done for cyber risk management.

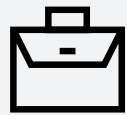

## The C-Suite View*:
### Improving Cost Effectiveness

**82%** said that cybersecurity spending will come under increased scrutiny and should align with identified risks to the business.

*The insight from interviews conducted with 56 organizations globally are reflected throughout the report as the "C-Suite View".

Cyber Crossroads
The Need for the Cyber Standard of Care

## Building a Foundation

The Cyber Crossroads foundational Cyber Standard of Care is inspired by the development of such standards in other disciplines — particularly medicine. Our goal is to provide a governance framework to help organizations manage their cyber risk. The Cyber Standard of Care as presented in this paper focuses on organizational cyber risk management processes that should occur regularly. The emphasis is on what a governance process should look like *before* an attack occurs. As we move ahead, there is work to be done regarding how to expand upon the principles expressed in the Cyber Standard of Care to address governance processes relevant to cyber incident responses following a cyber-attack.

This paper presents a step-by-step approach to ensuring that reasonable cybersecurity standards can be met by any organization, of any size, in any industry.

In addition, the Cyber Standard of Care we propose:
- Engages every department and stakeholder in the organization to identify cyber risks so that the process of cyber governance is **pervasive**.
- Facilitates cyber risk visibility across the organization so that business users and leadership are appropriately **informed**.
- Ensures that actions taken to mitigate cyber risks are **reasonable** and address relevant circumstances.
- Establishes **accountability** within each organization for cyber risk mitigation.
- Encourages **continuous improvement** throughout all divisions of an organization.
- Provides every board of directors with a governance model for **oversight and monitoring** of cyber risk management.

Adopting the Cyber Standard of Care will help organizations and their stakeholders:
- Become more effective at managing cyber risk.
- Improve the cost-effectiveness of their cyber risk management practices.
- Enhance accountability, and help reduce organizational liability.
- Reduce governance friction across organizations and their stakeholders (for example, regulators, vendors, and clients) when managing cyber risk.

Organizations trying to protect themselves from cyber-attacks are not lacking for advice. They are inundated with guidance from standard-setting bodies, government agencies, and regulators, and solicitations from a range of suppliers of cybersecurity products and services. The growing number and severity of cyber-attacks is fueling demand for these products and services, but there remain many questions concerning their overall effectiveness. It is becoming increasingly difficult for organizations and their stakeholders to feel confident that they are deploying resources efficiently.

The problem is exacerbated by the daunting scope of cyber risk. The US Council of Economic Advisers concluded that US firms lost 0.8% of their market value in the seven days following news of a breach (averaging $498 million per adverse event for the 186 firms in their study).

## Current State of Cyber Risk Management

Cybersecurity is a growing concern for organizations large and small, public and private, and across the information technology and operational technology domains. When organizations do not meet security expectations, potential consequences include direct financial losses, reputational damage, business interruption, reduced client and investor confidence, and financial penalties stemming from regulatory fines or lawsuits. A major cyber-attack will also typically demand a great deal of senior management's time and attention, imposing heavy opportunity costs. As a result, cybersecurity has become a top priority for directors and officers, who may be held accountable for failures, regardless of their lack of expertise.

A variety of costs can stem from a cyber-attack (see Figure 1). Their composition will depend on the nature and purpose of the attack. For example, ransomware attacks can generate sizeable business interruption expenses, whereas phishing attacks that target sensitive data may result in high forensics costs to identify and plug the breach, as well as expenses to notify customers, in compliance with applicable data breach laws.

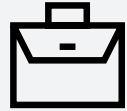

### The C-Suite View: Make Cyber Governance Pervasive

**85%** said accountability for cyber risks should be distributed throughout the organization.

**66%** said efforts should be made to educate the business on cyber risk.

**55%** said involving the business will help IT departments better identify and quantify exposures.

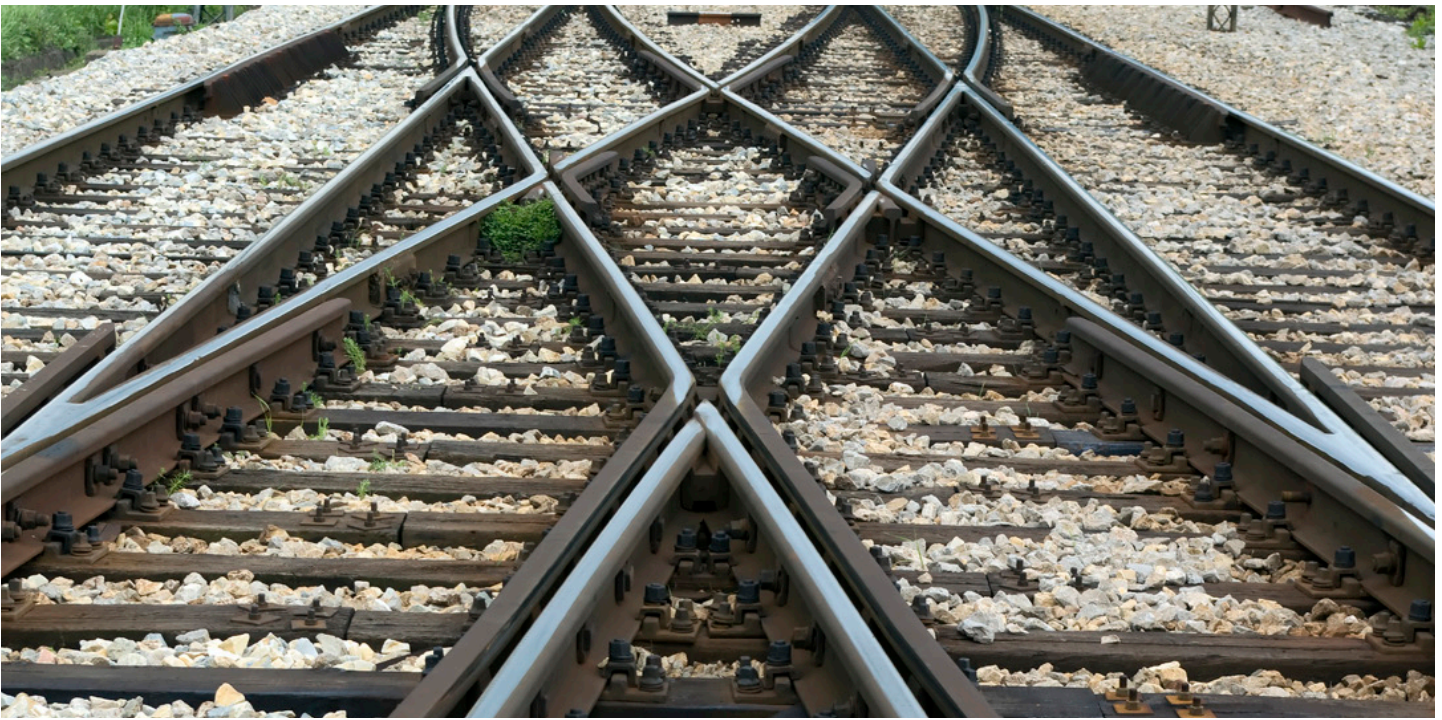

FIGURE 1
# How the Costs of a Cyber-Attack Mount
Cost categories, in order typically incurred

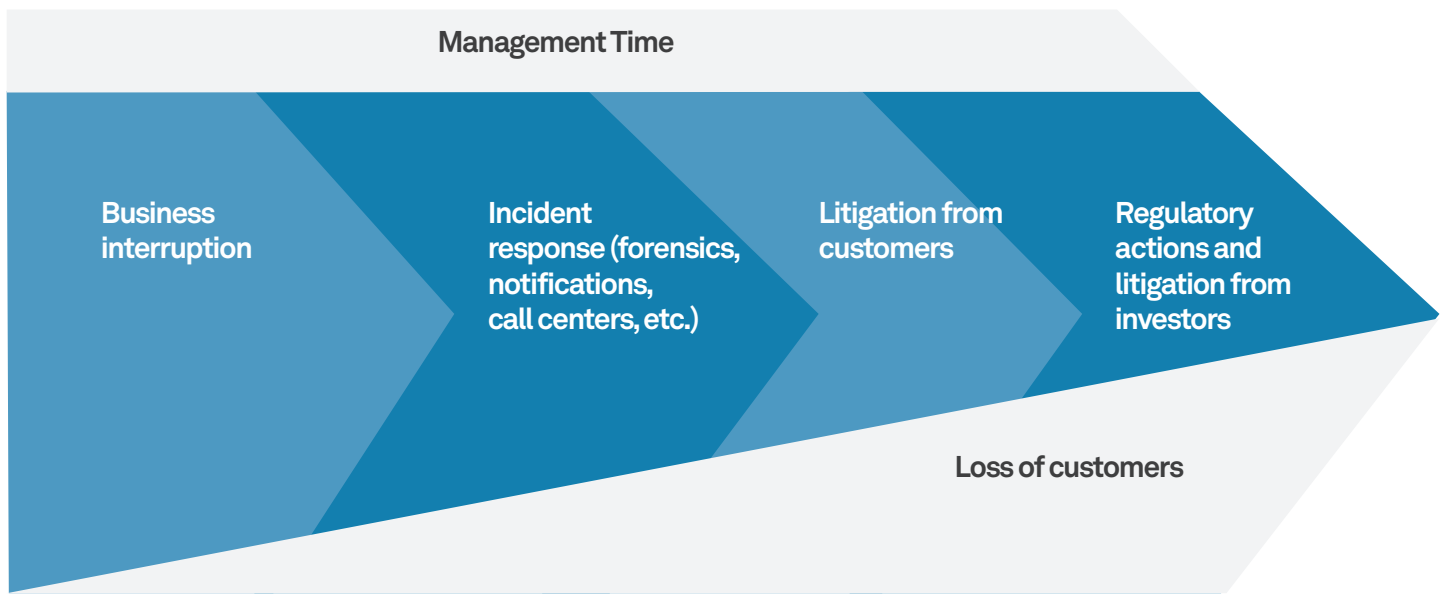

Despite considerable and growing expenditure on security products and services and the extensive guidance on managing cybersecurity risk offered by regulatory agencies and standards bodies, the number of successful cyber-attacks continues to rise (Beazley 2020). Many explanations have been offered for the apparent failure of these interventions to reduce the incidence and severity of cyber-attacks, ranging from organizations not purchasing the products and services appropriate to their specific needs, to the confusing nature of the different, yet overlapping, standards. At present, there is no agreed upon mechanism by which organizational leadership can demonstrate that the actions they are taking are both appropriate and adequate to reduce security risks given the organization's specific circumstances.

Large data breaches affecting many users naturally generate litigation on the part of at least some of the customers of the companies affected (see Figure 2). In the US, this litigation often takes the form of class action lawsuits, which at times garner large settlements (see Figure 3).

FIGURE 2

# Biggest Data Breaches of the 21st Century as Measured by Users Impacted

| Company Name | Company Activity | Users Impacted |
|---|---|---|
| Yahoo | Web services company | 3,000,000,000 |
| Sina Weibo | Microblogging | 538,000,000 |
| Marriott International | Hotel chains | 500,000,000 |
| Adult Friend Finder | Adult dating site | 402,200,000 |
| MySpace | Social media | 360,000,000 |
| NetEase | Provider of online mailbox services | 235,000,000 |
| Zynga | Online game producer | 218,000,000 |
| LinkedIn | Social network for business professionals | 165,000,000 |
| Dubsmash | Video messaging services | 162,000,000 |
| Adobe | Design software | 153,000,000 |
| MyFitnessPal | Fitness app owned by Under Armour | 150,000,000 |
| Equifax | Credit bureau | 147,900,000 |
| eBay | Online auctioneer | 145,000,000 |
| Canva | Graphic design tool company | 137,000,000 |
| Heartland Payment Systems | Payment card transaction processor | 134,000,000 |

Source: CSO Online, 2020

FIGURE 3

# Major Settlements for Class Action Lawsuits Triggered by Data Breaches

| Company Name | Settlement Amount (US$) |
|---|---|
| Anthem | $115 million |
| Target | $28.5 million ($18.5 million for states, $10 million for consumers) |
| Home Depot | $19.5 million |
| Sony (PlayStation network breach) | $15 million |
| Ashley Madison | $12.8 million ($11.6 million for consumers, $1.2 million for states and the Federal Trade Commission) |
| Sony (employee information breach) | $8 million |
| Stanford University Hospital and Clinics | $4.1 million |
| AvMed Inc. | $3.1 million |

Source: Morgan and Morgan, 2018

**To address these issues, many organizations are moving toward implementing a governance framework regarding the management of their cyber risk.**

However, these often maintain a traditional focus on investments in cybersecurity products and on decisions made by a chief information security officer (CISO). We believe the emphasis should be on broadening responsibility for cybersecurity throughout the organization, and on actions focused on processes and people, rather than purely technology.

The Cyber Standard of Care can help support a positive relationship between the CISO and other parts of the business. The CISO cannot do the job alone — for example, staff in front line operations need to ensure that they understand how a cyber event might impact on their day-to-day operations, procurement staff need to understand how cyber risk can be managed in the supply chain, and HR staff need to understand the important role that they have to play in ensuring that access to systems and data is only available to authorized personnel.

In addressing both technological and human factors, the Cyber Standard of Care will provide a template for those organizations looking to establish a state-of-the-art cyber risk management governance process, and will also allow for consistent risk governance benchmarking across organizations.

## The Role of Directors and Officers

Directors and officers are increasingly playing a central role in their organizations' cyber risk management, a responsibility historically reserved for the CISO. Given the scale of the damage that cyber risk can cause to an organization, management of this risk must be viewed as an enterprise-wide problem rather than simply an IT one.

Nevertheless, cyber risk remains alien territory for many outside of the IT and security departments. Although surveys place cyber risk near the top of boardroom concerns, board members often have a poor understanding of its complexity and scale (World Economic Forum 2019). Anecdotally, many boards struggle to provide effective oversight regarding cyber risk because they view it as a technical problem requiring a technical solution. In reality, cybersecurity is as much a governance issue as an operational one (Marsh and Microsoft 2018).

In many instances, the losses associated with cyber-attacks affect reputations and profits. As the financial stewards of a company, directors and officers are responsible to shareholders for these losses. Directors' and officers' potential liability for cyber risk has increased markedly as cybersecurity risk management has come to be seen as a governance responsibility. It is not easy to quantify the extent to which this has driven recent increases in the price of Directors' and officers' liability (D&O) insurance, but it is likely to have been a factor.

Organizations have faced, and will likely continue to face, litigation in connection with cybersecurity issues. Lawsuits against companies in all sectors — including finance, hospitality, technology, and more — have made headlines in recent years. While some lawsuits have yet to result in liability, even an unsuccessful action presents potential exposure in terms of defense costs, distraction for management and board members, and reputation. As boards' roles in cybersecurity risk management evolve, such lawsuits may also evolve, potentially becoming more successful. Clearly, the current litigation landscape suggests that organizational leaders need to be proactive in their approach to organizational cyber governance.

## Mounting Pressure and Requirements for Cybersecurity

In recent years, laws and regulations concerning cybersecurity and data privacy have proliferated. Furthermore, many countries have introduced new enforcement arms to oversee these regulations. For example, the United States Securities and Exchange Commission (SEC) has a cyber unit within its enforcement division to pursue cyber-related misconduct. An investigation conducted by the unit led in 2018 to Yahoo agreeing to pay a $35 million penalty to settle charges brought by the SEC that it failed to disclose in its public filings, for nearly two years, a major data breach. (Yahoo had only released information about the breach during the sale of its operating unit to Verizon).

The European Union's General Data Protection Regulation (GDPR), which came into force in 2018, has been at the forefront of global privacy regulation. Its scope is broad, extending to organizations that are not physically established in the EU but that offer goods and services to data subjects in the EU, or that monitor data subjects' behavior in the EU (Skadden 2018). Between May 25, 2018, when the regulation came into force, and November 30, 2019, 22 supervisory authorities in EU and European Economic Area states issued approximately 785 fines under the GDPR (EDPB 2020). The Information Commissioner's Office (ICO) in the UK levied the largest fine announced to date, against British Airways, which initially exceeded $200 million (Information Commissioner's Office 2019).

Beyond lawsuits, there is pressure to improve security in an effort to maintain customer confidence. A study published by IBM Security in 2019 that quantified the costs from data breaches across 16 countries, found that lost business — which comprised revenue losses, business disruption, system downtime, and new customer acquisition — was the largest of four major cost categories ensuing from a data breach, averaging 36% of total costs (IBM 2019.)[4]

The global nature of modern supply chains has also added to cybersecurity challenges. Organizations increasingly expect their vendors and others they do business with to have rigorous cybersecurity protocols in place, in part to avoid costly business interruption losses in the event of an attack.

## Existing Standards for Cyber Risk Management

Multiple sources of guidance exist to help organizations reduce their cyber risk. For example, reputable standard-setting bodies including the International Standards Organization (ISO) and the National Institute of Standards and Technology (NIST) have issued extensive recommendations.

Many security experts acknowledge standards such as NIST 800-53 as best practice. However, the standards landscape for cybersecurity is anything but straightforward. For some sectors, the landscape is so complicated that a "guide to the guidance" is required. For example with respect to financial services firms in the United States, the Financial Services Sector Coordinating Council (FSSCC) created a mapping tool to help organizations navigate the various requirements and directives (FSSCC 2020). (A summary of the leading cybersecurity guidance and standards is included in Appendix 2.)

Many organizations have adopted one or more standards in a "best effort" fashion. In doing so, they presumably hope these are the most appropriate measures for their risk exposure, but they cannot be sure. The costs of implementing guidance from standard-setting bodies can be high, and some have cited it as a major barrier to adoption for the NIST standard (Tenable 2016).

Cyber Crossroads describes a flexible management process and governance structure that can be overlaid on existing organizational activities in implementing cybersecurity technology and practices.

## Complex Organizational Structures Obscure Cybersecurity Accountability

The often-complex structure of large organizations complicates the challenge of achieving effective cybersecurity in the absence of an overarching standard of care. For example, the boards and management committees of subsidiary companies can play a critical role in determining and identifying reasonable actions that can decrease risk for the parent company. Subsidiary entities will generally need to consider local regulatory requirements and legislation specific to the jurisdictions in which they operate. Subsidiaries that operate largely independently of their parent companies face different risk profiles from those that make extensive use of group-level IT and other services.

As far as regulators are concerned, subsidiaries have the same responsibilities as independent companies. Regulators may therefore expect subsidiary entities to treat centralized group-level services in the same way that they would treat any third-party service provider. This has the potential to introduce a different type of dynamic into the group-subsidiary relationship. Enacting the Cyber Standard of Care will clarify where accountability lies, and how such accountability should be enforced across the different legal entities.

Cyber Crossroads
The Need for the Cyber Standard of Care

# Principles for the Cyber Standard of Care

The Cyber Crossroads team reviewed professional standards of care in other fields, including construction, manufacturing, and medicine. Among these, the medical field has a particularly strong emphasis on a process-oriented approach to setting a standard of care as a governance matter. No single document summarizes the medical standard of care; however, various court cases have shaped its development over time. Some of these cases are described below, along with the "success criteria" they helped to set.

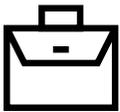

### The Medical Standard of Care as an Inspiration

Doctors and other medical personnel handle situations in which there is often some likelihood of a patient's death. The medical standard of care is not meant to replace the many guidelines, technologies, or standards in various medical sub-fields, nor does it prescribe how medical practitioners should care for their patients. Rather, it describes a responsible process for choosing *which* existing standards of practice to employ, as described in scientific journals and medical texts. The medical standard of care also does not describe success criteria. Rather, it defines a process for choosing among professional practices that, if followed, will help avoid failure (which can be characterized as "malpractice"). It therefore offers a set of criteria that medical workers can apply to the choice of treatment protocols, technologies, or vendors.

A crucial aspect of the medical standard of care is that medical practitioners are not legally judged in terms of patient outcome, be it life or death. Instead, they are accountable for whether the decision process they followed was sound.

Clearly, there are important differences between medicine and cybersecurity. However, the main elements of the medical standard of care that have been derived from court cases are instructive and can be applied to cyber risk management.[1]

[1] The court cases referenced in this report occurred in the US and UK, but the concept of a medical standard of care is certainly not limited to those countries. Other countries have their own variations. For example, in February 2013, Germany enacted the Gesetz zur Verbesserung der Rechte von Patientinnen und Patienten (Patients' Rights Act), which embodies many similar concepts to those found in US and UK law.

# Components of a Cyber Standard of Care

Our analysis of court cases (referenced below) that have helped define a professional standard of medical care identified four key components that can be incorporated into the Cyber Standard of Care. These are: circumstances, discovery, actions, and outcomes.

## Circumstances

There are circumstances that aggravate risk. These will derive from an organization's business practices and its external operating environment. It is critical to identify the circumstances that give rise to risk and identify appropriate risk management responses.

This component derives from the 1957 UK case of *Bolam vs. Friern Hospital Management Committee*. The court found that a medical practitioner falls below the minimum standard of care if they fail to take actions that are reasonable given the circumstances.

## Discovery

A qualified individual must be the one to determine the source of a problem or the immediate risk. This individual must have the proper background education, knowledge, and skill to assess the situation. Without firsthand, expert knowledge of the conditions, it is impossible to make an accurate assessment of risk.

This element of the medical standard of care derives from the 1964 UK ruling on *Hedley Byrne & Co. Ltd. v Heller & Partners Ltd*. In that case, the court introduced the notion of "reasonable reliance," which indicates that a person can reasonably rely on another for guidance if the person being relied upon possesses the requisite expert knowledge.

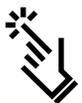

## Actions

Actions taken must be reasonable given the circumstances. Reasonableness is defined by whether peer practitioners would take similar action in similar circumstances, or if there is a case study or experiment that indicates the scientific validity of such an action.

This component derives from the 1985 US case *Hall v. Hilburn*. The courts determined that negligence is not present if others with similar expertise would have taken the same action given the circumstances.

## Outcomes

Actions have outcomes. Often the outcomes are intentional, but not always. Even if actions are reasonable, it is still possible to have unintended outcomes that are beyond an organization's control.

This component derives from the 2015 UK case *Montgomery v Lanarkshire Health Board*. This case requires medical practitioners to warn patients of potential unintended outcomes that may result from medical procedures, despite there being no fault on the part of the medical professionals performing such procedures.

# Elements of a Medical Standard of Care that are Relevant to the Cyber Standard of Care

| Circumstances giving rise to risk | Discovery | Actions taken to manage risk | Outcomes from actions taken |
|---|---|---|---|
| What in the organization's business practices or external operating environment exposes it to cyber risk? | Has a qualified individual evaluated the risk? | Are they reasonable? Would peer practitioners do the same? | Did the actions work as intended? Were there any unintended effects? Have lessons been documented and integrated into future practices? |

## Success Criteria to Meet a Standard of Care

Organizations need success criteria in order to assess their progress in implementing the standard of care. An organization should satisfy the following set of criteria regarding cyber governance before it can be considered to have met the Cyber Standard of Care.

**Pervasive:** An organization's cyber risk management process should engage the entire organization and its closely tied external partners. Each department contributes to, and can help to manage, the organization's aggregate cyber risk. Therefore, it is important for business users across all departments to identify and understand organizational cyber risks.

**Informed:** The entire organization must be aware of its current and future cyber risks and make decisions in line with a holistic perspective of the risks they face. Awareness and training exercises about the cyber risks that directly pertain to employees' operational roles can facilitate the education required to enable an informed organization.

**Reasonable:** Organizations must take actions that are directly relevant to the cyber risks that they face and that have been proven generally valid and effective. Validity can be established based on empirical evidence, where available, demonstrating the success of actions in mitigating given cyber risks. Alternatively, it can be demonstrated through peer organizations' experience employing the same actions over time. Actions must be proportionate to the risks the organization faces. For example, meeting relevant technical/operational standards could help to demonstrate reasonableness, but may not always be sufficient. Likewise, it may be possible to demonstrate reasonableness without meeting a particular standard in full.

**Accountable:** An organization should give its employees the knowledge and skills needed to participate in meeting the organization's cybersecurity goals. They also should give them specific, measurable targets to meet. Clear lines of responsibility are key to successful risk mitigation and remediation. There should be incentives or consequences to encourage and enforce accountability.

**Continuous Improvement:** The fact that an organization's threat landscape is constantly changing is a prime reason that cybersecurity is a continuing concern. Cyber risk management is not a one-time operation. Organizations should study their past cyber risk management successes and failures and incorporate their lessons into future training.

**Oversight and Monitoring:** The process of cyber risk management needs to be monitored carefully. This should be part of the responsibility of senior management in the organization. Investments in cybersecurity must be managed purposefully, and adjusted based on evolving organizational security needs.

## The C-Suite View: Scope for Improvement in Assessing Cyber Exposures

**69%** said that determining organizations cyber risk exposure requires a business context beyond what is accessible to the CISO and security team.

# Implementing the Cyber Standard of Care

In the spirit of the medical standard of care and the previously described success criteria, following is a sample implementation of the Cyber Standard of Care. The proposed implementation, which can be completed as a desktop exercise, can be highly beneficial in managing cyber risk.

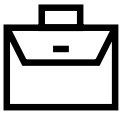

We created the sample implementation after interviewing organizations from different industries about their existing cyber risk governance process (see Figure 4). Their collective processes were subsequently mapped to the Cyber Standard of Care principles.

Organizations should implement the Cyber Standard of Care on a regular timetable aligned to enterprise-wide business processes. The sample discussed below follows a quarterly cadence.

The sample implementation uses a hierarchical approach to employing the Cyber Standard of Care, which may not be consistent with how some organizations operate. Each organization determines its business hierarchy, governance and risk management structure, and roles and responsibilities of its senior management based on its business characteristics and the challenges it is facing, such as digitalization. Each organization should have its unique structure when it implements the Cyber Standard of Care. The maturity of each organization, culture, or even regional preferences may also result in the differences in risk management structure. Therefore, we encourage organizations to adapt the Cyber Standard of Care in such a way that its implementation would be consistent with their existing process and governance style.

The Cyber Standard of Care builds on the components identified in the medical standard of care. Organizations cannot afford to wait decades for a standard to evolve through court cases, as occurred in the medical field. The Cyber Standard of Care can be implemented swiftly. The four main elements comprising the standard — circumstances, discovery, actions, and outcomes — are reflected in the process.

FIGURE 4
# Elements of the Cyber Standard of Care

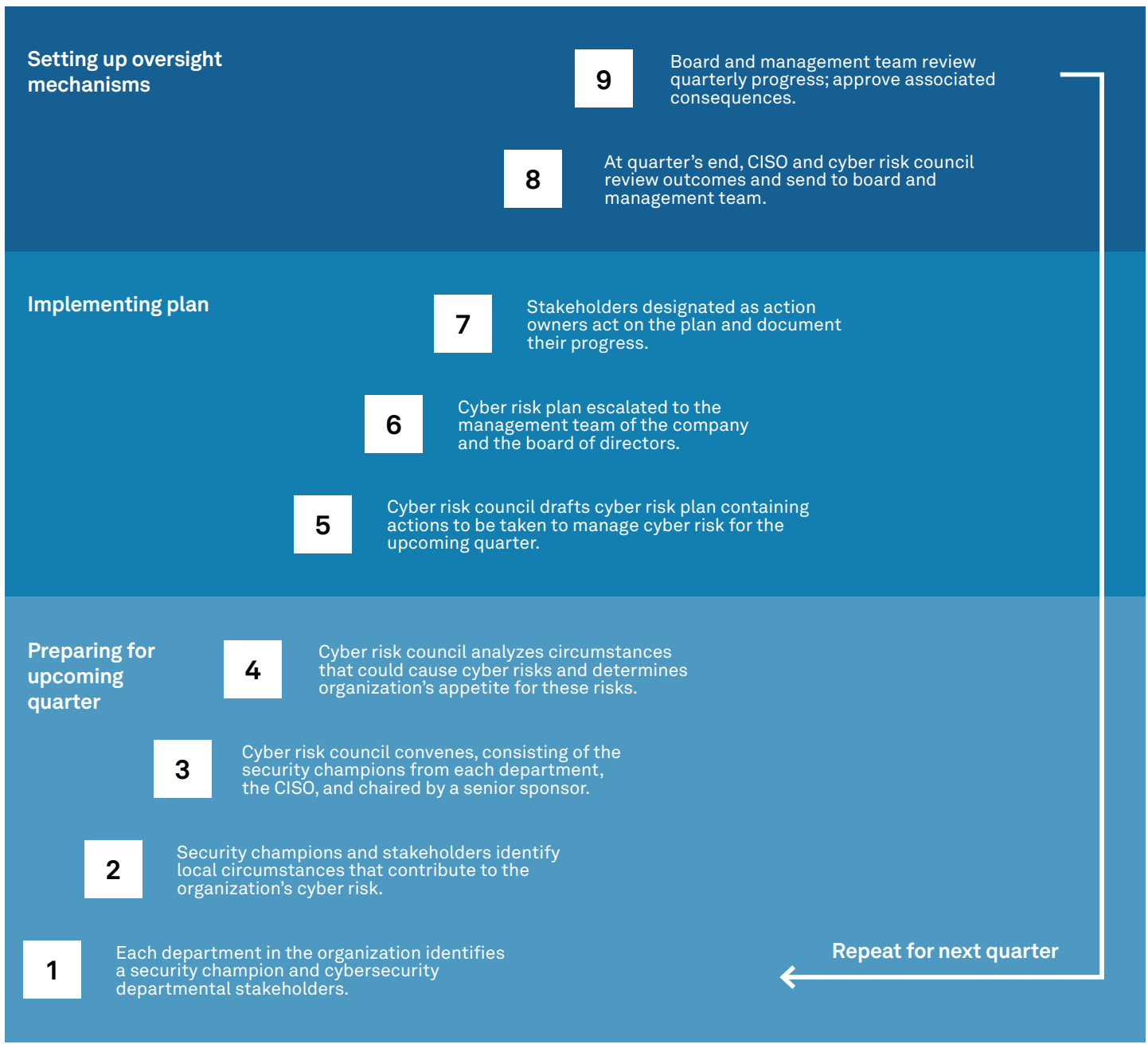

**Setting up oversight mechanisms**

9 — Board and management team review quarterly progress; approve associated consequences.

8 — At quarter's end, CISO and cyber risk council review outcomes and send to board and management team.

**Implementing plan**

7 — Stakeholders designated as action owners act on the plan and document their progress.

6 — Cyber risk plan escalated to the management team of the company and the board of directors.

5 — Cyber risk council drafts cyber risk plan containing actions to be taken to manage cyber risk for the upcoming quarter.

**Preparing for upcoming quarter**

4 — Cyber risk council analyzes circumstances that could cause cyber risks and determines organization's appetite for these risks.

3 — Cyber risk council convenes, consisting of the security champions from each department, the CISO, and chaired by a senior sponsor.

2 — Security champions and stakeholders identify local circumstances that contribute to the organization's cyber risk.

1 — Each department in the organization identifies a security champion and cybersecurity departmental stakeholders.

**Repeat for next quarter**

# Sample Cyber Standard of Care

The steps outlined in Figure 4 are described in more detail below.

**Step 1**
At the beginning of each quarter, each department in the organization identifies a cybersecurity champion and departmental stakeholders involved in making operational decisions. A cybersecurity champion should be knowledgeable about the span of operations across the department, hold relationships with stakeholders across the department's functions, and have an interest in cybersecurity. Designating departmental champions and stakeholders will aid in the discovery process for identifying cyber risks in the organization. Because these stakeholders are involved in departmental decisions, they should be qualified to identify the cyber risks to the organization that result from these decisions.

*Success Criterion Addressed: Pervasive*

**Step 2**
Security champions and stakeholders from across the organization identify local circumstances that contribute to the organization's cyber risk. These might relate to activities undertaken by the department that could increase cyber risk for the rest of the organization (for example, developing a new application or service, entering a new business partnership, or taking action that could create disgruntled employees). We refer to these as "risks from" the department's activities. Alternatively, circumstances might relate to heightened risk exposure to the department's business activities due to factors unique to that department — we refer to these as "risks to" the department's activities. Cyber Crossroads has established a list of questions that departmental stakeholders should consider as they identify potential "risks from" and "risks to" the business (see page 21).

*Success Criteria Addressed: Pervasive and Informed*

**Step 3**
After establishing local circumstances for each department, the organization should create a cyber risk council consisting of the security champions from each department, the CISO, and chaired by a senior sponsor in the organization. The cyber risk council's responsibility is to review all circumstances that the stakeholders and security champions have documented for the quarter. Upon convening, the cyber risk council should consider additional circumstances that the stakeholders have not identified, including imminent future risks. These could include circumstances beyond the purview of any single department, such as changes in the evolving political climate or the emergence of previously unsuspected attack vectors.

The cyber risk council should not operate in a cyber risk silo, and should consult other individuals in the organization who are familiar with its cyber risk, including those who lead the organization's risk management, insurance, compliance, and internal audit functions.

*Success Criterion Addressed: Informed*

**Step 4**
The primary role of the cyber risk council after reviewing and documenting additional risk circumstances is to determine the organization's cyber risk appetite. Clearly, a "boil the ocean" approach to protecting the organization against every conceivable cyber threat is not feasible: the cyber risk council must make choices grounded in their knowledge of the organization's operations. Documenting these priorities will form the core of the cyber risk plan established for the quarter.

*Success Criteria Addressed: Informed and Reasonable*

**Step 5**
After determining the circumstances that need to be addressed based on risk appetite, the cyber risk council must determine what actions should be taken for the quarter. This is the critical point in the decision-making cycle. Skill and judgement need to be applied to ensure that the actions proposed are appropriate to the business internal and external operating context and directly related to minimizing the risk for the circumstances that were selected as priorities. The key term is "reasonable," which means that the actions proposed need to be validated to reduce the cyber risk, for example by evidence-based studies or anecdotally from peers. Each action should be assigned an owner, and accountability for delivery should be directly embedded within the established mechanisms that the organization uses to make sure that things get done. Actions should be specific and measurable, there should be consequences on action owners should delivery fail, and line management should be responsible for ensuring that these consequences are applied. These measures will help to drive accountability for the actions. The actions, owners, and consequences will be added to the cyber risk plan for the quarter. Cyber Crossroads has established a list of questions to help the cyber risk council evaluate reasonable actions for given circumstances (see page 22).

*Success Criterion Addressed: Reasonable*

**Step 6**
After the cyber risk council agrees on its draft of the cyber risk plan, they will escalate it to the company's management team and board of directors. The CISO will present to the board and a representative of the board will agree to the plan. The CISO will then turn over the relevant parts of the plan to the security champions and relevant managers in accordance with the accountability mechanisms used within the company. This step demonstrates that there is appropriate oversight and monitoring by the directors and officers.

*Success Criteria Addressed: Oversight and Monitoring*

**Step 7**
The stakeholders designated as the action plan owners will document their progress toward defined targets and report in a centralized fashion to the security champion and the CISO. When an owner completes an action, they will mark it as such. The owner will also document outcomes of the action taken, which includes any potential positive results, for example, attacks foiled, cyber events, or side effects related to the action.

*Success Criteria Addressed: Pervasive, Informed, and Accountable*

**Step 8**
At the end of each quarter, the CISO and the relevant stakeholder's managers will review the outcomes. Those actions that were not completed will be evaluated to determine if consequences should be levied on the security stakeholder. This will be at the discretion of the relevant managers in accordance with the accountability mechanisms used within the company. The CISO will collate a report of the actions taken and present the results to the board of directors and management team.

*Success Criteria Addressed: Reasonable and Accountable*

**Step 9**
The board of directors and management team will agree to the quarterly progress report delivered by the CISO and approve any associated consequences. At this stage, the CISO will make recommendations for priority key risk indicators (KRIs) or circumstances that were not addressed to focus on for next quarter's planning cycle. Budgeting and investment in the governance process will be managed and adjusted accordingly based on the past cycle. Cyber Crossroads has established a list of questions for the board and management team to determine whether the Cyber Standard of Care was met (see page 23). The process described is not unlike that of sales forecasting and planning.

*Success Criteria Addressed: Continuous Improvement and Oversight and Monitoring*

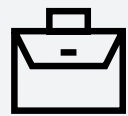

# The C-Suite View:
## Need for a Cyber Risk Council, More Planning

**96%** support the establishment of a cyber risk council to oversee cyber risk.

**80%** say cyber risk planning should take place more than once a year.

# The Spectrum of Cyber Risk Governance

In the course of our interviews with senior executives, it became clear that cyber risk governance means different things to different organizations. Some organizations have just begun their cyber risk governance journey. Others are highly advanced and mature; we believe they are already meeting a cyber standard equivalent to what we propose.

The following table compares those only beginning their cyber risk governance journey to those that meet the Cyber Standard of Care across the success criteria we have identified. The "low" and "high" maturity descriptions reflect responses from interviewed organizations.

## Maturity Across the Cyber Standard of Care Spectrum

| Success Criterion | Low Maturity | High Maturity |
| --- | --- | --- |
| Pervasive | Organization centralizes responsibility for identifying cyber risk issues in the CISO's office. | The planning process is distributed throughout the organization and is inherently pervasive because the entire organization is involved in the discovery process. During discovery, each department assigns a security champion, and engages employees to identify the cyber risks that are a result of departmental operations and decisions. |
| Informed | Mandatory security and compliance training is the only mechanism to ensure all stakeholders are involved in cyber risk management. | The planning process charges departmental stakeholders and security champions with determining how their department's business functions contribute to cyber risk. The act of documenting circumstances ensures that everyone is engaged in identifying risks and organizational circumstances that contribute to cyber risk. This makes for informed employees at all levels of the organization and fosters a culture of cyber security across the organization. |
| Reasonable | Actions taken to manage cyber risk are largely reactive to current threats and do not look ahead to future threats. | The planning process requires directly linking actions to established circumstances that contribute to cyber risk. This ensures that actions are relevant to the pressing organizational cyber risk issues. The cyber security council selected these actions because they have been demonstrably effective or are consistent with their peers' recommendations. This process helps to ensure the reasonable use of resources to address cyber challenges. |
| Accountable | The primary party accountable for all cyber risk is the CISO and their IT/Security team. | Planning broadens accountability through the development and documentation of a quarterly plan. Senior management and the board approve the plan, immediately instilling some accountability at the organizational leadership level. The plan defines "action owners" across all departments. Each action has success criteria, and consequences, if actions are not implemented. By documenting cyber risk management efforts in a leadership-approved plan and establishing owners for each action and consequences for inaction, accountability becomes distributed across the organization. |
| Continuous Improvement | There are no clear mechanisms for drawing lessons from the organization's experience with cyber risks or for adjusting future behavior. | The quarterly planning process supports continuous improvement, with planned actions changing based on newly identified circumstances. When actions are reviewed, the organization is encouraged to learn from experience and adjust future behavior accordingly. The documentation of circumstances each quarter, review by the cyber risk council, and agreement to the plan facilitates learning across the organization. |
| Oversight and Monitoring | Boards of directors do not have the same insight into cyber risk as they have into other types of organizational risks. | Directors and officers oversee and monitor the organization's cyber risk management efforts. There is active engagement of leadership. As the quarter concludes, leadership reviews the actions taken and makes decisions concerning where to direct risk mitigation efforts. The process embodies how and why the organization enacts cyber standards and industry best practices, consistent with how boards oversee other major risks. |

# The Cyber Standard of Care Journey

Given the need for various groups across an organization to engage in developing a cyber risk governance process, Cyber Crossroads developed a series of questions to help make progress toward meeting the criteria. Each set of questions targets a different line of defense within the organization, including departmental stakeholders, the cyber risk council, and the board of directors. These question sets should help organizations navigate the journey towards the Cyber Standard of Care.

Depending on the organization's maturity, some questions will be more relevant than others. Some aspects of an organization's maturity may be a function of cultural differences across geographies. Organizations that find themselves on the "low maturity" spectrum may find the question set for the departmental stakeholders most immediately relevant, as compared with those that find themselves at a "high maturity" level.

## The Cyber Standard of Care Includes Four Components:

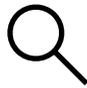 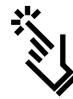

**Circumstances**

Identifying the unique contextual factors that contribute to cyber risk for an organization.

**Discovery**

Appointing qualified assessors to pinpoint the precise cyber risks throughout the organization.

**Actions**

Specifying risk mitigation procedures that directly address the organization's unique circumstances.

**Outcomes**

Reviewing outcomes and sharing lessons, while acknowledging that adverse outcomes can occur even in the best-managed organizations.

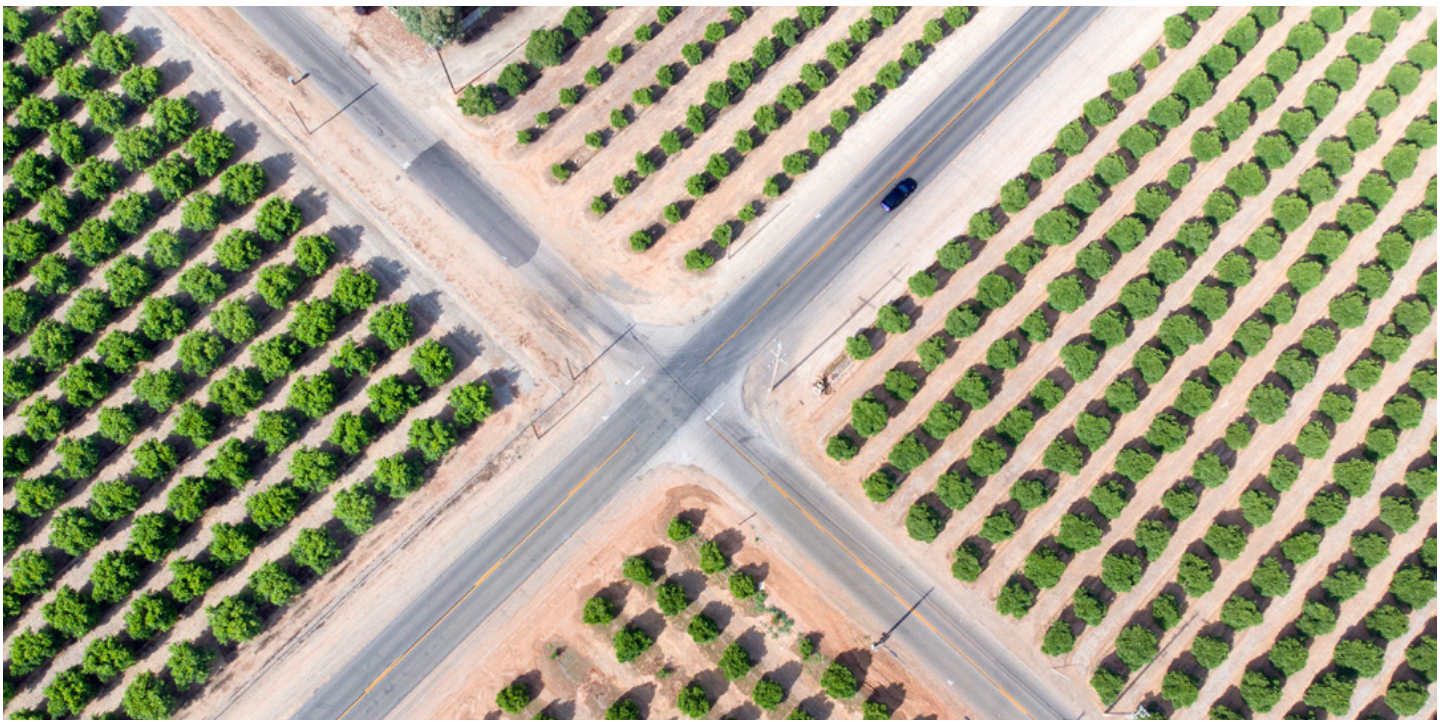

# Question List for Departmental Stakeholders

As noted in step two of the proposed implementation of the Cyber Standard of Care, departmental stakeholders will need to enumerate the possible operating circumstances that could generate serious cyber risks for their organization. Departmental stakeholders can use the following questions as a starting point.

### Departmental Stakeholders' List of Questions

| | |
|---|---|
| What are the potential cyber risks to our departmental business operations? | • What sensitive data sets do we hold? Have there been additions in the past quarter?<br>• Who do we share sensitive data with (internally and externally)?<br>• What are our critical services, and what is the nature of their dependence on IT systems? How would a sustained outage affect these services? Have these vulnerabilities increased in the past quarter?<br>• Who are our critical third-party service providers? What would the impact be if there were a sustained interruption in their services? To what extent does their cybersecurity impact our own? |
| What potential cyber risks does the organization face arising from our departmental business operations? | • What activities are we undertaking that could attract adverse external attention to the department or to the rest of the organization?<br>• Are we operating in any particularly high-risk geographies?<br>• Is there any reason why we might be facing a heightened risk from disgruntled internal employees (or contractors)?<br>• What IT applications do we have that are unique to the department, and do these present contagion risk to other organizational infrastructure? |
| How well are we managing cyber risk within the department? | • Do we have a cyber champion and are they equipped to perform this role?<br>• Have all of the circumstances that can contribute to cyber risk been accurately recorded and monitored? Are we complying with all relevant organizational IT/security policies? Where there are exceptions, have these been authorized and how are we mitigating residual risks?<br>• Are we achieving the right levels of cyber assurance from our third-party service providers?<br>• Do we have crisis response plans in place, and have we tested them?<br>• Are our specific departmental requirements adequately addressed in the organization-level cyber plan? If not, what additional local mitigations are required? |

# Question List for Cyber Risk Councils

As noted in step five of the proposed implementation of the Cyber Standard of Care, a cyber risk council will be responsible for determining if the actions selected are appropriate and reasonable for their circumstances and for regulatory requirements. The following questions can be used as a starting point for the cyber risk council.

## Cyber Risk Council's List of Questions

| | |
|---|---|
| Are we meeting the minimum standards required? | • Are there specific regulatory standards that we need to meet?<br>• Are we compliant with the requirements of our customers?<br>• Are we compliant with the requirements of our insurance policies? |
| Are we confident that we have accurately captured our organization's circumstances? | • Have departments provided credible descriptions of the risks that relate to their business operations?<br>• Are there any significant changes to the organization's wider operating circumstances (including major P&L issues, new products and services, new markets, and M&A) that could affect our cyber risk profile?<br>• Are there any significant changes to the external threat environment, now or in the immediate future?<br>• Have any significant incidents occurred to others in our sector, or beyond, that should cause us to reconsider our actions and priorities? |
| Is our implementation appropriate to our circumstances? | • Is the implementation of each of our key controls comparable to good practice within our sector?<br>• Are we able to provide a logical rationale for our choices?<br>• Do we face either exceptionally high risk exposure (or exceptional low risk tolerance) relative to our sector? What additional measures might be expected of us given these exceptions?<br>• Are we able to demonstrate that we have responded effectively to changes in our internal or external circumstances? |
| Are we making the right scheduling and trade-off decisions? | • Where are we outside our comfort zone for cyber risk?<br>• Can we demonstrate that the sequencing of our mitigation steps is optimized against our risk reduction targets? |
| Are our actions in line with the priorities and risk appetite that have been set by the board? | • Have we communicated the rationale for our decisions effectively to the board?<br>• Does the board understand the extent to which our actions have been effective, and do they appreciate the residual risks that we are running?<br>• Have we provided the board with the assurance they need that our discovery and decision-making processes are fit for purpose? |

## Question List for the Board of Directors

As noted in step nine of the proposed implementation of the Cyber Standard of Care, the board of directors will be responsible for determining if their organization's cyber risk management efforts meet the standard by comparing their process to the success criteria described. The following questions can be used as a starting point for the board.

### Board of Directors' List of Questions

| | |
|---|---|
| Pervasive | • Are all the departments across the organization engaged in the discovery of cyber risk? |
| Informed | • How are business users across the organization equipped to identify circumstances contributing to cyber risk? |
| Reasonable | • How are actions defined to specifically address the circumstances identified? |
| Accountable | • How well implemented is the plan of action? Does it have the right targets, owners and incentives? |
| Continuous Improvement | • How are departments/employees learning from each other and building capacity and expertise from past circumstances, actions, plans, and incidents? |
| Oversight and Monitoring | • How is the organization investing in the governance of its cybersecurity program/spending to provide both assurance and justification? |

# Benefits of the Cyber Standard of Care

Cyber Crossroads recommends the Cyber Standard of Care as a means to overcome current deficiencies in the cybersecurity governance ecosystem. There are four major benefits of enacting the Cyber Standard of Care:

### 1. Improve the Effectiveness of Cyber Risk Management.

The Cyber Standard of Care, as proposed, will improve the effectiveness of cyber risk governance processes. Specifically, it will improve visibility of cyber risks across the organization. Further, the Cyber Standard of Care will provide organizations a reporting structure they can consistently follow and escalate to organizational leadership. The information collected as part of the quarterly plan will empower stakeholders to take actions — such as selecting cyber tools or services — that are directly tied to their circumstances. Also, evidence of precautions taken will be accessible to external regulators or insurers that are part of the organization's risk management ecosystem.

### 2. Improve the Cost Management of Cyber Risk Investments.

The Cyber Standard of Care necessitates a structured approach to enumerating risks and acting on them in a logical fashion. While there will be cost associated with implementing a robust cyber governance process, there will be offsetting financial benefits. Organizations are making large investments in cybersecurity: it would not be unreasonable to spend 5% of a cybersecurity budget on governance to ensure the other 95% is spent efficiently. A cyber governance process that requires actions to directly link to circumstances also provides a clearer picture for external parties to assess an organization's responsible spending on cyber risk.

Cyber insurance coverage has proven to be a critical protection for organizations of all sizes, and premiums have been rising. Clear documentation concerning a thoughtful process in selecting cyber risk solutions could help to better position an organization as an insurance risk and potentially help garner better results in negotiating insurance protection than might otherwise be available. Insurance is only one aspect of an organization's total cost of risk. Self-insured retentions and losses that exceed the policy's limits also contribute to the cost of cyber risk. If through the Cyber Standard of Care, organizations can reduce, or eliminate, the impact of cyber-attacks, these costs could also fall.

## 3. Improve Accountability and Reduce Liability.

The Cyber Standard of Care will help organizations allocate accountability for cybersecurity appropriately. This is particularly pertinent at large organizations where subsidiaries operate in widely varying regulatory environments. In these organizations, relationships with local regulators are likely to be enhanced by a coherent, group-wide approach to cyber risk management that emphasizes local accountability.

Additionally, the liability of directors and officers could potentially be reduced should the Cyber Standard of Care be successfully implemented. Implementing the Cyber Standard of Care, with its extensive audit trail as described in the sample implementation, could serve to demonstrate that the organization took "due care" to mitigate cyber risk and could aid in the defense of suits brought against the directors and officers in the event of a cyber incident.

## 4. Reduce Governance Friction.

The Cyber Standard of Care will not only benefit organizations themselves, but also their regulators and insurers. The Cyber Standard of Care provides a common language and approach to implementing a wide range of options available to manage cyber risk. Because the Cyber Standard of Care does not specify what needs to be done at the technical or operational level, it is jurisdictionally agnostic. It specifies a process-oriented approach that can build on, and be adapted to satisfy, the guidance provided by authorities and standards organizations in different locales. The value of a single overarching framework for exercising oversight of cyber risk management should not be underestimated. The framework could serve to consistently report cyber risk management processes to clients, and could equally be requested of vendors. Since many companies now operate on a global scale, and the insurance industry underwrites organizational risk spanning multiple countries, there is a need for a common understanding of how the directors and officers of organizations will be expected to perform their oversight functions. The process approach embodied in the Cyber Standard of Care fills that void and thereby reduces friction otherwise present when managing cyber risk.

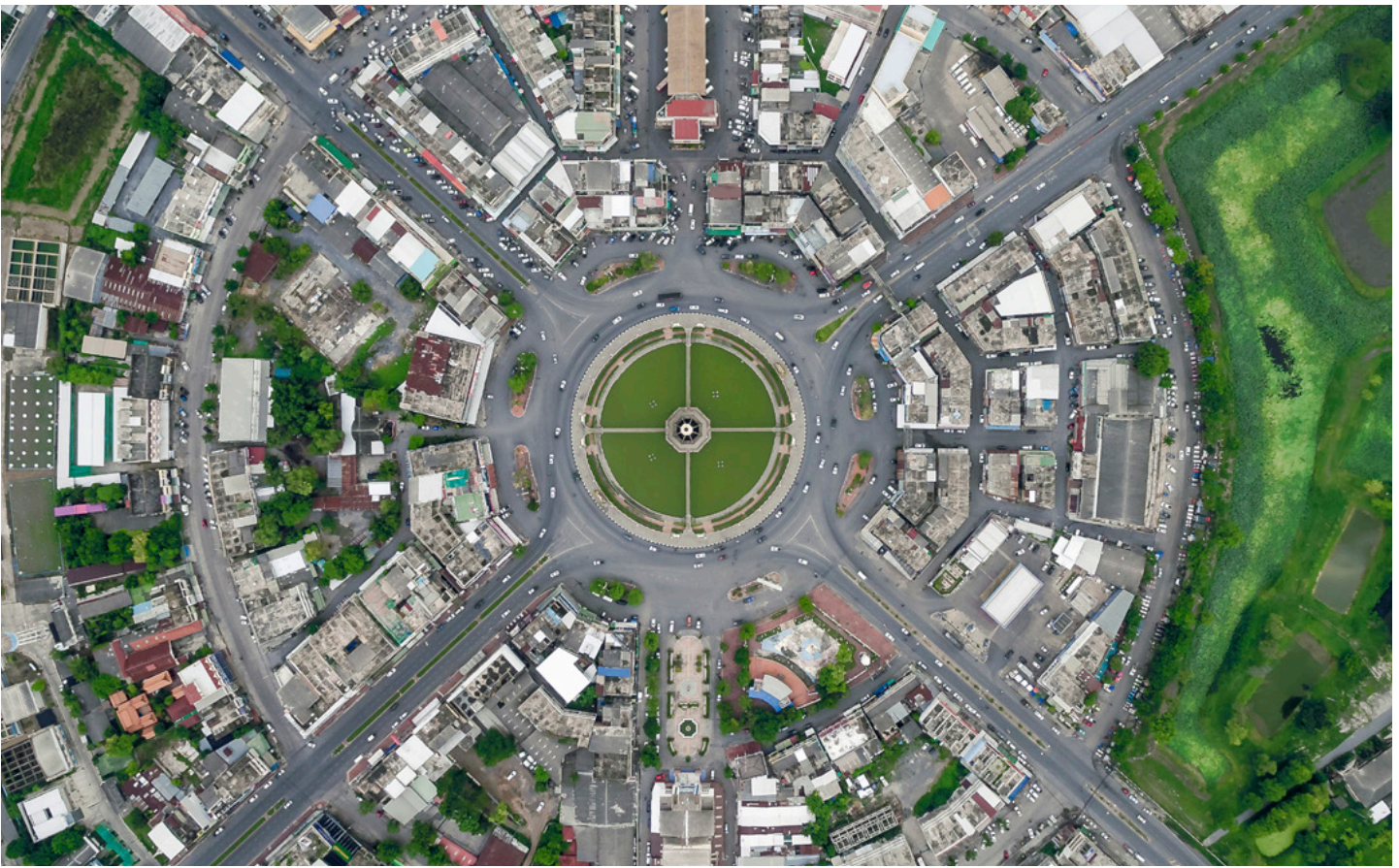

# Conclusion

The management of many of the risks that organizations confront has, over the years, become increasingly codified, sometimes in legislation.

For example, Congress passed the Sarbanes-Oxley Act following Enron's collapse in 2002, requiring companies to implement processes to protect against fraud, with directors and officers accountable for implementation. Groups such as the US Department of Homeland Security Solarium Commission have suggested expanding SOX to include the governance of cyber risk (Solarium 2020). The Cyber Standard of Care, however, could do the job without legislative action and could prove a better solution.

Processes and procedures, not just technical standards, need to be spelled out, something that standard-setting bodies are not well equipped to do. The Cyber Standard of Care would take the guesswork out of organizational efforts to make smart, cost-effective investments in cybersecurity. It would facilitate the creation of a cybersecurity culture in organizations by infusing cyber risk thinking into normal business operations and processes. It could also assist organizations and their directors and officers in rebutting subsequent challenges to the adequacy of cyber governance efforts, whether by regulators or litigants.

# Appendices

## Appendix 1
## Methodology

The Cyber Crossroads team approached this undertaking with the intention of establishing a set of guiding principles for the Cyber Standard of Care and a roadmap to implement it. To establish the Cyber Standard of Care principles, a literature review was conducted across existing cyber standards, guidelines, and practices (see Appendix 3) and across the medical standard of care. From the existing literature, a draft set of principles was created, consisting of two parts — the components of a standard of care and success criteria for the standard of care. The Cyber Crossroads team then reviewed the principles with financial and insurance regulators who provided feedback.

Regulators suggested that providing general principles for a Cyber Standard of Care would not be sufficient. To facilitate adoption, it was recommended that a sample prescriptive implementation be provided to organizations.

The Cyber Crossroads team then interviewed senior executives actively engaged in cyber risk management at over at 56 industry leading organizations globally to understand their current cyber risk governance processes. Interviewees included CISOs, chief risk officers (CROs), general counsels, and chief operating officers (COOs). Industries represented in the sample included consumer goods, energy, financial services, health care, media/entertainment, professional services, retail, technology, hospitality, and utilities.

These interviews enabled the Cyber Crossroads team to compare existing processes to the principles we had identified for the Cyber Standard of Care. Combining both the governance processes from the interviews and the principles, the Cyber Crossroads team established a sample implementation of the Cyber Standard of Care. The sample implementation is intended to help organizations determine how one may approach enacting the Cyber Standard of Care.

## Appendix 2
## Board Resources for Cyber Risk

A wide variety of publications offer board-level support on issues of cybersecurity. These range from targeted reports (Grant Thornton 2019), to standards (BS 31111:2018), guidance (NCSC UK 2019, NCSC NZ 2019), and frameworks (FERMA and ECIIA 2018, WEF 2017). Given its relative novelty, the sources of guidance available to boards are more variable and less mature than for topics that have been boardroom agenda items for far longer. A brief overview of the main existing support resources is described below.

These boardroom resources are strategically helpful, orienting boards towards general best practices and the importance of cyber risk for the organization. They are lacking in proposing a clear governance system that could be employed to help achieve control over the organization's cyber risk. The Cyber Crossroads Cyber Standard of Care helps to fill this gap.

### World Economic Forum (WEF) Report on Advancing Cyber Resilience

Through this publication, the WEF (2017) introduced a boardroom-oriented cybersecurity framework and a range of "toolkits." The document cites a perceived demand for board-level support and tools on this issue, based on primary data collected by WEF. In response to this demand, the 10 principles of the framework are introduced which serve as an expression of broader corporate governance objectives such as accountability, transparency, and awareness. More specifically, the framework addresses the allocation of board-level responsibility for cybersecurity, the availability of required capabilities, the integration of cyber resilience in the broader business strategy, and the development of a strategy for planning, reviewing, and improving the cybersecurity performance of the organization. Each principle is associated with a series of proposed questions (toolkits) designed to enable self-assessment and nurture the necessary conversations between governance and management functions. The document also includes a number of references to notable cybersecurity frameworks, while also integrating the concepts of resilience and risk in a cybersecurity context. This includes the introduction of risk matrices and broader risk management principles, while also referencing the WEF's previously introduced Cyber Risk Framework (WEF 2017:18).

### The United Kingdom and New Zealand's National Cyber Security Centers (NCSC UK 2019, NCSC NZ 2019)

The importance of board-level support on issues of cyber risk has been the focus of efforts from national bodies, like the UK's and New Zealand's National Cyber Security Centers (NCSC UK 2019, NCSC NZ 2019). As a result, the former has developed the Cyber Security Toolkit for Boards, characterized by its accessible approach and tone. It attempts to provide a high-level overview of cybersecurity governance, while also introducing tangible actions at a board-level and at an organizational level. This approach is reflected structurally, with each of its central themes including actions, and a framing of 'what good looks like' through a series of self-assessment questions. The guidance covers a series of themes that range from approaches to cybersecurity expertise growth, to cyber risk management, collaboration, and incident response planning. Furthermore, the content leverages NCSC UK's portfolio of resources and available guidance, while also referencing the (limited) cybersecurity regulation applicable to its core national audience. NCSC NZ's (2019) board and senior executive guidance comes in the form of its "Charting Your Course: Cyber Security Governance" framework. The output employs a six-stage approach to cyber resilience, which, similar to the previous examples, addresses culture, roles, risk management, collaboration, cyber program development, and resilience measurement. Each stage introduces a series of associated concepts and is accompanied by specific guidance. Broader cyber governance actions, structures, roles and responsibilities, which can span beyond the board of directors, are also recommended.

### Association Publications

At a European level, the Federation of European Risk Management (FERMA) and the European Confederation of Internal Auditing (ECIIA) (2018) have addressed the intersection between corporate governance and cybersecurity in their proposed cyber risk governance model entitled "At the Junction of Corporate Governance & Cybersecurity" (FERMA & ECIIA 2018). The output employs a conception of cyber risk based on an integration of the Organization for Economic Cooperation and Development (OECD) Digital Security Risk Management principles in their *Digital Security Risk Management for Economic and Social Prosperity* and FERMA and ECIIA's "Three Lines of Defense" model as described in their report entitled *Guidance on the 8th EU Company Law Directive* (OECD 2015, FERMA & ECIIA 2010). The former is consistent with similar documents and provides guidance on a set of common cyber risk issues. These range from social factors, such as awareness, skills and cooperation, to more technical/operational topics such as risk assessments, security measures, and resilience. In contrast, the Three Lines of Defense model defines domains of responsibility, which place holistic oversight of cyber risk at a board level. The resulting cyber governance model includes a prescriptive, risk-centric overview of relevant organizational functions (IT, data management, HR), actors (risk managers, data protection officer, CISO, finance officer, and compliance officer), audit activities, and relationships with external stakeholders (such as insurers, regulators, and vendors). Guidance is also provided on the interaction between risk managers and internal auditors. As part of its appendices, the document includes a comparative overview of the Network and Information Security Directive and General Data Protection Regulation — the two key European Union regulatory initiatives that interact with and impact organizational cybersecurity (European Union Agency for Cybersecurity 2016, European Union 2016).

### BS 31111:2018: "Cyber Risk and Resilience — Guidance for the Governing Body and Executive Management"

As one of the first standards to explicitly focus on cyber risk and resilience from a governance/executive management perspective, British Standard 31111:2018 provides a broadly applicable base of non-technical guidance. It attempts to promote the development of cyber resilience by aligning the focus and understanding of executive management on a series of key principles. These principles are derived from "other management standards or frameworks," and broadly relate to organizational resilience, risk management and information security. More specifically, they are structured around four central themes (clauses): Building Cyber Resilience: Core Principles; The Organizational Foundations of Cyber Risk and Resilience; Building Cyber Risk Management and Resilience Capability; and Assessing the Resilience of the Organization. Each is accompanied by sub-themes, which outline a series of recommended actions and responsibilities for executive management. The document also includes a series of informative appendices that offer guidance on improving assurance and governance, and understanding cyber culture. In line with other similar resources, the standard also includes a series of self-assessment questions addressed to corporate governance actors that reflect its central themes.

<p style="text-align:center">***</p>

In summary, the importance of board-specific resources on cybersecurity is increasingly recognized by a number of leading bodies. This is reflected in the growing range of publications that seek to complement traditional, high-granularity, prescriptive cybersecurity frameworks by focusing on corporate governance support. These have evolved from offering a localized, functional view of cybersecurity towards a holistic, organizational perspective. That said, none suggest a clear process that directors and officers can oversee or a mechanism for evaluating the efficacy of the cyber risk management program in place. Such a process could be covered by the Cyber Standard of Care.

# Cybersecurity Frameworks

### The NIST Cybersecurity Framework

Since its initial release by the National Institute of Standards and Technology (NIST) in 2014, the NIST Framework for Improving Critical Infrastructure Cybersecurity, has gathered widespread support. The framework is comprised of three dimensions:

1. A "core" that defines activities, outcomes, and references based on five cybersecurity "functions" — identify, protect, detect, respond, recover.

2. A four-tier structure that attempts to facilitate the qualitative self-assessment of an organization's cyber risk management efforts.

3. A "profile" component that aligns the framework core with the business context and enables contrasting current and target states.

The framework's latest version (1.1, NIST 2018) supplements the original structure with a series of incremental changes — most notably, an increased emphasis on cyber supply chain risk management. Overall, the framework is designed to support and structure cybersecurity efforts and resources in a manner that reflects the organization's environment, baseline capabilities and development aspirations. While only implicit in v.1.1, "Governance and Enterprise Risk Management" is identified as one of the high-priority development areas within the framework's roadmap (NIST 2019a).

As an initial measure in this direction, NIST adapted its Baldridge Excellence Framework to create the Baldridge Cybersecurity Excellence Builder (NIST 2019b) — a self-assessment tool designed to better understand an organization's cybersecurity (risk) performance in its operational and strategic context.

The Cyber Security Framework (CSF) is also central to the second revision of NIST SP 800-37 "Risk Management Framework for Information Systems and Organizations" (NIST 2019c), suggesting a stronger integration between the CSF and NIST's SP 800 Series. However, its recommended resources and references include other compatible frameworks such as ISO27K and COBIT, discussed below.

### The ISO/IEC 27K Family of Standards

Through its breadth, the ISO/IEC 27K family of standards addresses a variety of dimensions of IT security. With guidelines pertaining to Information Security Risk Management (ISO/IEC 27005:2018), Governance of Information Security (ISO/IEC 27014:2013) and Guidelines for Cybersecurity (ISO/IEC 27032:2012), the main standard for which certification can be provided is ISO/IEC 27001:2013 (Information Security Management Systems, or ISMS). Its normative requirements integrate information security management and risk management, and outline a framework for establishing, evaluating, and improving a Security Management System based on the organization's context. In line with the terminology proposed in ISO/IEC 27000:2018, the ISMS standard prescribes an explicit role for "top management," which is defined based on organizational direction-setting and control, exemplified through C-suite roles. It is also accompanied by a series of prescriptive controls, which are further expanded in ISO/IEC27002:2013. Finally, the structure, principles and approaches employed throughout the family of standards are consistent with and refer to the broader ISO portfolio (for example, Quality Management Systems — ISO 9000:2015, Risk Management — ISO 31000:2018, and Corporate Governance — ISO 37000/38500). Thus, despite a limited immediate orientation towards supporting board-level engagement with cybersecurity, some valuable guidance is provided.

### COBIT 2019 Enterprise Governance of Information and Technology Framework

The COBIT 2019 Enterprise Governance of Information and Technology framework from the Information Systems Audit and Control Association (ISACA) addresses information security in the broader context of IT Governance, with a primary focus on "value creation." Its core model is structured around "domains" that differentiate governance and management objectives. In this explicitly prescriptive model, 40 objectives are proposed in association with each of the domains (COBIT 2019b:17). The framework also introduces a series of governance system design factors, each accompanied by a normative outline of its recommended dimensions.

Despite its prescriptive depth, emphasis on holism, and relatively distinct approach to information security as a secondary dimension of wider IT governance, the COBIT framework adopts compatibility with other standards (including ISO and NIST) as a core objective. As a result, each of its suggested components is accompanied by references to "related guidance," comprising references to relevant standards, regulatory requirements and other frameworks. Where appropriate, the COBIT framework also proposes example metrics, process overviews based on a range of capability-based activity models, recommended actor involvement based on organizational roles, as well as potential dependencies in information flows and items — such as inputs and outputs from/to other components (COBIT 2019a, COBIT 2019b).

## Other Frameworks

Two other noteworthy inclusions in the cybersecurity framework landscape, based on their recognition and adoption rates, are the Center for Internet Security (CIS) Controls (2019), and the Payment Card Industry's Data Security Standard (2018). The former aims to provide a systematic approach for prioritizing security actions based on the organization's profile. It proposes 20 security controls, categorized as basic, foundational or organizational, each with a series of sub-controls that are prescribed based on the "implementation groups" model. Descriptions, rationales, and entity relationship diagrams are also provided for each of the main controls, resulting in a general, robust, yet functionally confined prescriptive model of cybersecurity actions, with little reference to or explicit support for corporate governance. In contrast, PCI DSS (2018) is a technical standard aimed at all actors that are either involved in payment card processing or interact with cardholder data. It provides a certification system, a range of implementation best practices, and an explicit set of requirements, guidance and testing procedures. Despite its relatively limited scope, PCI DSS introduces a thorough prescriptive data security model that, while centered on payment card data, can have broader effects on an organization's cybersecurity stance.

**Appendix 3 References**

# Authors/Reviewers

**Cyber Crossroads Founder/Director:
Arvind Parthasarathi**

Arvind Parthasarathi is the Founder and Director of Cyber Crossroads, which he conceptualized when discussions on cyber risk with executives around the world kept coming back to the same themes on governance. He was previously the Founder and CEO of Cyence (now merged with NYSE: GWRE), which created a cyber risk analytics platform to quantify the financial impact of cybersecurity risks. Before that, he was President/CEO of Yarcdata (now merged with NYSE: HPE), which created a platform for data discovery. He was previously Senior Vice President and General Manager of Informatica (Nasdaq: INFA).

**Professor Paul Cornish,
London School of Economics**

Professor Cornish is Visiting Professor at LSE IDEAS, the foreign policy think tank at the London School of Economics. He has advised several national cybersecurity strategies, contributed to the International Telecommunication Union Guide to Developing a National Cybersecurity Strategy, and was instrumental in the establishment of the Oceania Cyber Security Centre in Melbourne, Australia. Previously, he was Co-Director of the Cyber Security Capacity Centre at Oxford University and Professorial Fellow in Cyber Security at the Australian National University.

**Professor Madeline Carr,
University College London**

Madeline Carr is Professor of Engineering Science at University College, London, and the Director of the UK-wide Research Institute in Sociotechnical Cyber Security (RISCS), which looks at the human and organizational factors of cybersecurity. She is also the Director of the Digital Technologies Policy Lab, which supports policymaking to adapt to the pace of change in society's integration of digital technologies.

**Cyber Crossroads Principal Investigator:
Dr. Gregory Falco, Stanford University**

Gregory Falco is a Cyber Risk Research Fellow at Stanford University's Cyber Policy Center, where his research focuses on cyber risk governance and management. He is also a Research Affiliate at MIT's Computer Science and Artificial Intelligence Laboratory and a Cyber Research Fellow at Harvard University's Belfer Center for Science and Technology. Prior to his PhD in Cybersecurity at MIT, he co-founded and led Accenture's Smart City Strategy business.

**Professor Sadie Creese,
Oxford University**

Sadie Creese is Professor of Cyber Security in the Department of Computer Science at the University of Oxford and a member of the faculty of the Blavatnik School Executive Public Leaders Programme, where she lectures on cybersecurity topics. She is the founding Director of the Global Cyber Security Capacity Centre (GCSCC) at the Oxford Martin School, where she continues to serve as a Director, conducting research into what constitutes national cybersecurity capacity. She leads the Oxford team's collaboration with the World Economic Forum's Shaping the Future of Cybersecurity and Digital Trust Platform. She is also Co-Chair of the Lloyds Register Foundation sponsored Foresight review of Cyber Security for the Industrial Internet of Things (IIoT).

**Dr. Myriam Dunn Cavelty,
ETH/Swiss Federal Institute of Technology**

Myriam Dunn Cavelty is a Senior Lecturer in Security Studies and Deputy for Research and Teaching at the Center for Security Studies. She was a Visiting Fellow at the Watson Institute for International Studies at Brown University and a Fellow at the Stiftung Neue Verantwortung in Berlin, Germany. She advises governments, international institutions, and companies in the areas of cybersecurity, cyber warfare, critical infrastructure protection, risk analysis, and strategic foresight.

**Professor Claudia Eckert,
Technical University of Munich**

Claudia Eckert is Professor of IT Security at the Technical University of Munich. Her research focuses on new concepts, methods, and technologies to increase security and trustworthiness of IT-based systems and applications. She is also Director of the Fraunhofer Institute for Applied Integrated Security, which supports firms from all industries and service sectors in securing their systems, infrastructures, products, and offerings.

**Professor Gen Goto,
University of Tokyo**

Gen Goto is Professor of Law at the University of Tokyo Graduate Schools for Law and Politics. He was a visiting Professor at the National University of Singapore and Harvard Law School and is a Director of the Japanese Law and Economics Association. His main field of research is comparative corporate governance, with a recent focus on the role of shareholders in the Japanese context.

**Andrew Grotto,
Stanford University**

Andrew Grotto is Director of the Stanford Cyber Policy Center's Program on Geopolitics, Technology, and Governance. He is a William J. Perry International Security Fellow at the Freeman Spogli Institute for International Studies and a Visiting Fellow at the Hoover Institution, both at Stanford University. Previously, he was the Senior Director for Cybersecurity Policy on the National Security Council at the White House, in both the Obama and Trump Administrations.

**Sean Kanuck,
Stanford University**

Sean Kanuck is an Affiliate at the Center for International Security and Cooperation at Stanford University. Previously he served as the first US National Intelligence Officer for Cyber Issues from 2011 to 2016, after a decade with the CIA's Information Operations Center and the White House National Security Council. Previously, he was Director for Cyber, Space & Future Conflict at the International Institute for Strategic Studies in London, Distinguished Visiting Fellow at Nanyang Technological University in Singapore, and Distinguished Fellow with the Observer Research Foundation in India. He is CEO of Exedec, a cyber strategy advisory firm that advises financial sector clients on the future of information technology and associated risks.

**Dr. Herbert Lin,
Stanford University**

Herbert Lin is Senior Research Scholar for cyber policy and security at the Center for International Security and Cooperation and Hank J. Holland Fellow in Cyber Policy and Security at the Hoover Institution, both at Stanford University. He is also Chief Scientist, Emeritus for the Computer Science and Telecommunications Board, National Research Council (NRC). His research interests focus on the policy and national security dimensions of cybersecurity and cyberspace. In 2016, he served on President Obama's Commission on Enhancing National Cybersecurity.

**Dr. Jamie Saunders,
Oxford University/University College London**

Jamie Saunders is Fellow at the Oxford Martin School at Oxford University and a Visiting Professor at University College London. He served for 29 years in UK Government including GCHQ and the Cabinet Office. Most recently, he was Director of the National Cyber Crime Unit, part of the UK's National Crime Agency and previously Director of International Cyber Policy at the UK Foreign and Commonwealth Office.

**Dr. Howard Shrobe,
Massachusetts Institute of Technology**

Howard Shrobe is a Principal Research Scientist at the Massachusetts Institute of Technology's Computer Science and Artificial Intelligence Laboratory (CSAIL). He served twice as the Defense Advanced Research Projects Agency (DARPA) program manager, leading programs in cybersecurity and software engineering. He has served as an Associate Director of CSAIL and of the MIT Artificial Intelligence Laboratory and was Vice President of Technology at Symbolics Inc. Dr. Shrobe is a thought leader in developing trustworthy, assured, and explainable systems and AI for government agencies and defense organizations.

**Prof. Larry Susskind,
Massachusetts Institute of Technology**

Lawrence Susskind is Ford Professor of Urban and Environmental Planning at the Massachusetts Institute of Technology, where he heads the Science Impact Collaborative and is Director of the Cyber Urban Critical Infrastructure program, where students are trained in matters of urban critical infrastructure security and have the opportunity to directly help critical infrastructure organizations and municipalities. Professor Susskind is the author or co-author of twenty books in eight languages and in 2020 was the winner of MIT's Digital Technology Teaching Award.

# Sponsors

**Cyber Crossroads thanks the following organizations for their sponsorship of the research.**

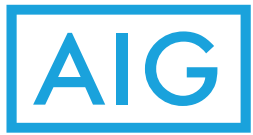
www.aig.com

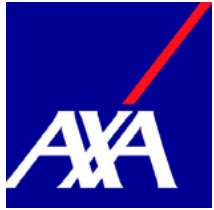
www.axaxl.com

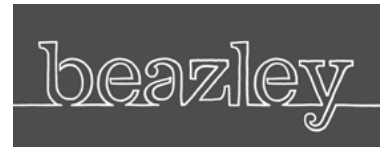
www.beazley.com

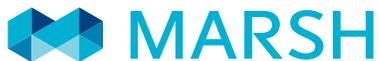
www.marsh.com

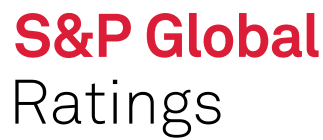
www.spglobal.com/ratings

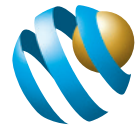
www.tokiomarine.com

In order to maintain the independence and integrity of Cyber Crossroads, sponsors were not involved in the research process or outputs. Sponsors did, however, provide directional guidance and feedback to ensure the findings were relevant and the recommendations were practical.